\documentclass[11pt,a4paper]{article}
\pdfoutput=1
\usepackage{jcappub, slashed}

\usepackage{bm,graphicx,amssymb}

\begin{document}

\title{Constraints on decaying dark matter from the extragalactic
gamma-ray background}
\author[a]{Shin'ichiro Ando}
\author[b]{and Koji Ishiwata}

\affiliation[a]{GRAPPA Institute, University of Amsterdam, 1098 XH
  Amsterdam, The Netherlands}
\affiliation[b]{Deutsches Elektronen-Synchrotron DESY, Notkestrasse
  85, 22607 Hamburg, Germany}

\emailAdd{s.ando@uva.nl}
\emailAdd{koji.ishiwata@desy.de}

\abstract{If dark matter is unstable and the mass is within GeV--TeV
  regime, its decays produce high-energy photons that give
  contribution to the extragalactic gamma-ray background (EGRB).  We
  constrain dark matter decay by analyzing the 50-month EGRB data
  measured with Fermi satellite, for different decay channels
  motivated with several supersymmetric scenarios featuring $R$-parity
  violation.  We adopt the latest astrophysical models for various
  source classes such as active galactic nuclei and star-forming
  galaxies, and take associated uncertainties properly into account.
  The lower limits for the lifetime are very stringent for a wide
  range of dark matter mass, excluding the lifetime shorter than
  $10^{28}$~s for mass between a few hundred GeV and $\sim$1~TeV,
  e.g., for $b\bar{b}$ decay channel. Furthermore, most dark matter
  models that explain the anomalous positron excess are also excluded.
  These constraints are robust, being little dependent on
  astrophysical uncertainties, unlike other probes such as Galactic
  positrons or anti-protons.}  \maketitle
 
\section{Introduction}
\label{sec:intro}    
\setcounter{equation}{0} 

Recent cosmological measurements of the cosmic microwave background
(CMB)~\cite{Hinshaw:2012aka,Ade:2013zuv} conclusively show that about
84\% of matter in the Universe is made of dark matter.  Evidently
these dark matter particles are electromagnetically neutral and also
stable compared to cosmological time scale.  However, it does not
necessarily mean that they have infinite lifetime.  In most cases,
dark matter is simply assumed to be stable, e.g., by introducing $Z_2$
symmetry in the Lagrangian.  One of the most popular examples of such
symmetries is $R$-parity in supersymmetric models, which is often
introduced to make protons stable.  It also stabilizes the lightest
superparticle (LSP) to be a candidate for dark matter.  However,
whether dark matter is absolutely stable or not, and hence whether
such symmetry as $R$-parity exists or is broken, should be tested in
light of modern data.

Dark matter decays produce cosmic ray particles, and details of the
final state depends on its interaction with standard-model particles.
In supersymmetry, $R$-parity violation (RPV) allows the LSP to decay
to standard-model particles, and this is indeed favorable for some
supersymmetric scenarios from cosmology arguments.  For instance, if
gravitino, a superpartner of the graviton, is the LSP, the next-LSP
may decay during big-bang nucleosynthesis, which destroys light
elements.  In RPV scenario, however, the gravitino dark matter with
mass of order 0.1--1~TeV makes it possible to overcome such
cosmological problem~\cite{Takayama:2000uz}. For gravitino dark matter
models, it was pointed out that the decay channels of gauge/Higgs
bosons plus lepton dominate the total decay width under bi-linear RPV,
which can produce observable high energy hadrons and
leptons~\cite{Ishiwata:2008cu}.  Another possibility is axino --- a
superpartner of axion, which in some supersymmetric scenarios becomes
the LSP~\cite{Rajagopal:1990yx,Goto:1991gq,
  Asaka:1998ns,Abe:2001cg,Choi:2006za,Choi:2011xt} and plays the role
of dark matter~\cite{Chun:1992zk,Covi:1999ty,Nakamura:2008ey} (see
also Refs.\,\cite{Banks:2002sd,Kawasaki:2007mk} for cosmology of
supersymmetric axion model).  Recently
Ref.\,\cite{Ishiwata:2014cra,Ishiwata:2013waa} studied the axino dark
matter with RPV in order to explain the observed baryon asymmetry.  In
this case RPV is crucial for baryogenesis, which is directly connected
to hadronic products in axino decay.\footnote{This is an explicit
  realization of the baryogenesis, which is proposed in
  Ref.\,\cite{Cheung:2013hza}.}  Finally, anomalous cosmic ray
positron fraction observed with
PAMELA~\cite{Adriani:2008zr,Adriani:2013uda},
Fermi~\cite{FermiLAT:2011ab} and
AMS-02~\cite{Aguilar:2013qda,Accardo:2014lma} motivated intensive (in
many cases phenomenological) discussions on leptophilic models, where
dark matter particles decay (or annihilate) mainly into multiple
leptons.

In this paper, we study the impact of dark matter decays on the latest
gamma-ray observations with the Large Area Telescope (LAT) onboard the
Fermi satellite.  In order to obtain stringent, yet robust,
constraints, we adopt the latest data for the extragalactic gamma-ray
background (EGRB)~\cite{Ackermann:2014usa}, which is isotropic
gamma-ray radiation from every direction of the sky. Strictly
speaking, EGRB is approximately isotropic since it partly consists of
gamma rays from point sources (we will see in detail later). However,
it is a good approximation to say that EGRB is isotropic
statistically.  Since we compare the EGRB data with the contribution
from dark matter decays that is the {\it statistical average} of
gamma-ray radiation from each of dark matter structures, this yields
results that are largely independent of any uncertain inputs such as
density profiles of dark matter halos.  In addition, we take into
account secondary gamma rays due to the inverse Compton (IC)
scattering by the charged particles from decays off the CMB photons as
discussed in Ref.\,\cite{Ishiwata:2009dk}.\footnote{IC photon in
  annihilating dark matter is studied in Ref.\,\cite{Profumo:2009uf}.}
This again yields no theoretical uncertainties if a decaying dark
matter model is specified; we do not include any other seed photons
such as regular optical emission from galaxies.  It is also known that
hadronic decay of dark matter is strictly constrained by measurements
of cosmic ray anti-protons~\cite{Adriani:2010rc, Ibarra:2008qg,
  Ishiwata:2010am, Cirelli:2013hv, Garny:2012vt}.  Besides the fact
that these anti-proton constraints are subject to much larger
uncertainties due to Galactic cosmic ray propagation, we will show
that the latest EGRB measurement gives more stringent constraints at
this point.  This should of course be revisited after the release of
AMS-02 data for anti-protons and other light elements.

Compared with the earlier work (see, e.g.,
Refs.\,\cite{Cirelli:2012ut, Cirelli:2009dv} and references therein)
based on the Fermi-LAT EGRB data collected for 10
months~\cite{Abdo:2010nz}, we improve upon the following points.
Firstly, we adopt the latest EGRB data measured with Fermi-LAT for 50
months~\cite{Ackermann:2014usa}.\footnote{Constraint on annihilating
  dark matter in light of the 50-month data is studies in
  Refs.\,\cite{Ajello:2015mfa,Ackermann:2015gga,DiMauro:2015tfa}.  }
Secondly, we address the systematic uncertainties that come from the
modeling and subtraction of the Galactic foreground emission, and find
that they reflect on uncertainties for lifetime constraints by a
factor of a few.  Lastly, in addition to purely phenomenological
power-law modeling of other components, we use more realistic
astrophysical components modeled individually with latest gamma-ray
luminosity functions calibrated with other wavebands.  We also adopt
the uncertainties associated with these modeling as prior information
in Bayesian statistical analyses to obtain lifetime constraints of
dark matter.  It is found that the obtained lower limit on the decay
lifetime is improved by about an order of magnitude compared with
Ref.\,\cite{Cirelli:2012ut}, for all the hadronic and leptonic decay
channels.  For example, for a conventional model of decays into
$b\bar{b}$, we find that the lifetime shorter than $10^{28}$~s is
excluded for dark matter masses between a few 100~GeV to 1~TeV.  In
addition, it is found that the parameter space which is motivated to
explain the positron excess is almost excluded.

This paper is organized as follows.  In Sec.~\ref{sec:dm}, we give
formulation of the EGRB computation from dark matter decays, and
several models that give finite lifetime to dark matter.
Section~\ref{sec:astro} then discusses three astrophysical components
that are considered for the analysis.  Results are presented in
Sec.~\ref{sec:result} along with details of the analysis, and we
discuss several important implications in Sec.~\ref{sec:discussion}.
Throughout this paper, we use cosmological parameters given by Planck
collaboration combined with WMAP~\cite{Ade:2013zuv}: present Hubble
rate, and density divided by critical density of matter, dark matter,
dark energy are $H_0=67.04~{\rm km~s}^{-1}\,{\rm Mpc}^{-1}$ (i.e.,
$h=0.6704$), $\Omega_{m}=0.3183$, $\Omega_{\rm dm}=0.2678$,
$\Omega_{\Lambda}=0.6817$, respectively. The critical density is
obtained by $\rho_c=1.054\times 10^{-5}\,h^2~{\rm GeV\,cm}^{-3}$.

\section{Decaying dark matter}
\label{sec:dm}
\setcounter{equation}{0}

Gamma rays from decaying dark matter is determined with three
ingredients: dark matter mass, lifetime, and energy distribution of
particles per decay, i.e.,
\begin{eqnarray}
m_{\rm dm}\, ,\  \tau_{\rm dm}\, , \ 
 \frac{dN_I}{dE}\ \  (I=\gamma,\, e^{\pm}, \cdots) \ ,
\end{eqnarray}
where the subscript `dm' stands for dark matter.  First two are free
parameters, while the $dN_I/dE$ is computed for each dark matter
model.  We formulate the EGRB from decaying dark matter based on
Refs.\,\cite{Ishiwata:2008cu,Ishiwata:2009dk}, then discuss dark matter
models on which we focus.

\subsection{Formulation}
\label{sec:formulation}
Dark matter contribution to the EGRB consists of two components:
primary and secondary gamma rays.  Secondary gamma rays are produced
by IC scattering off the CMB photons due to electrons and positrons
from dark matter decay.  The total gamma-ray intensity
$\Phi_\gamma^{\rm dm}(E_\gamma)$ is expressed as
\begin{eqnarray}
\Phi_{\gamma}^{\rm dm}(E_{\gamma})=
\frac{c}{4\pi}\frac{\Omega_{\rm dm}\rho_c}{m_{\rm dm}\tau_{\rm dm}} 
\int dt\, Q_{\gamma}^{\rm dm}(E_{\gamma},E'_{\gamma})\ ,
\end{eqnarray}
where $c$ is speed of light, $t$ is cosmic time,
$E_\gamma'=(1+z)E_{\gamma}$ is the energy of $\gamma$-rays when they are
produced at redshift $z$, and
\begin{eqnarray}
Q_{\gamma}^{\rm dm}(E_{\gamma},E'_{\gamma}) = e^{-\tau(z,E_{\gamma})} (1+z)
\left[{\cal P}_{\rm prim}(E'_{\gamma})+{\cal P}_{\rm ic}(E'_{\gamma})
\right]\ ,
\end{eqnarray}
where $\tau(z,E_\gamma)$ is optical depth, which we adopt the result
given in Ref.\,\cite{Gilmore:2011ks}. ${\cal P}_{\rm
  prim}(E'_{\gamma})$ and ${\cal P}_{\rm ic}(E'_{\gamma})$ are energy
distributions of primary and IC $\gamma$-rays, which are given by
\begin{align}
&{\cal P}_{\rm prim}(E'_{\gamma})=
\frac{dN_{\gamma}}{dE}(E'_\gamma)\ ,
\\
&{\cal P}_{\rm ic}(E'_{\gamma})=
\frac{c}{1+z} \int d E_e\, d E_{\gamma_{\rm CMB}} 
  \frac{d\sigma_{\rm IC}}{dE'_\gamma}(E'_\gamma,E_e,E_{\gamma_{\rm CMB}})
  f_{\rm CMB} (E_{\gamma_{\rm CMB}})
  \frac{Y_{e}(E_e)}{b_{\rm ic}(E_e)}\ ,
\end{align}
respectively, where $d\sigma_{\rm IC}/dE'_\gamma$ is the differential
cross section of the IC process, $f_{\rm CMB}$ is the spectrum of the
CMB (per unit volume per unit energy), and $b_{\rm ic}= (4/3)
\sigma_{\rm T} (E_e/m_e)^2 \rho_{\rm CMB}^{\rm (now)}$ is the energy
loss rate of the $e^\pm$ due to the IC process. Here $\sigma_T$ is
Thomson scattering cross section and $\rho_{\rm CMB}^{\rm (now)}\simeq
0.260\,{\rm eV\,cm}^{-3}$, and $Y_e(E_e)$ is defined by using
$dN_{e^{\pm}}/dE$ as
\begin{eqnarray}
Y_e(E_e) = \int^{\infty}_{E_e}\,  dE
\left[\frac{dN_{e^+}}{dE}(E) + \frac{dN_{e^-}}{dE}(E)
\right]\ .
\end{eqnarray}
Then after changing variable from $t$ to $E_\gamma'$, it is
straightforward to obtain
\begin{align}
E_\gamma^2\Phi_{\gamma}^{\rm dm}(E_{\gamma})&=
\frac{c}{4\pi}\frac{\Omega_{\rm dm}\rho_c}{m_{\rm dm}\tau_{\rm dm}} 
\frac{E_{\gamma}}{ H_0} 
\int^{\infty}_{E_\gamma} dE'_{\gamma}\frac{E_\gamma}{E'_{\gamma}}
\frac{Q_\gamma^{\rm dm}(E_\gamma,E'_{\gamma})}
{\sqrt{\Omega_{\Lambda}+\Omega_{m}(E'_\gamma/E_\gamma)^3}}
\nonumber \\
&=1.4 \times 10^{-7}~{\rm GeV \, cm}^{-2}\,{\rm s}^{-1}\,{\rm sr}^{-1}
\left(\frac{1\,{\rm TeV}}{m_{\rm dm}}\right)
\left(\frac{10^{27}\,{\rm s}}{\tau_{\rm dm}}\right)
\left(\frac{E_\gamma}{100\,{\rm GeV}}\right)
\nonumber \\
&\times \int^{\infty}_{E_\gamma} dE'_{\gamma}\frac{E_\gamma}{E'_{\gamma}}
\frac{Q_\gamma^{\rm dm}(E_\gamma,E'_{\gamma})}
 {\sqrt{\Omega_{\Lambda}+\Omega_{m}(E'_\gamma/E_\gamma)^3}}\ .
 \label{eq:intensity}
\end{align}

\subsection{Dark matter models}
\label{sec:DMmodel}
Decaying dark matter models can be classified into three categories by
the products after cascade decay: (i) lepton, (ii) hadron\,$+$\,lepton
and (iii) hadron.  Decaying dark matter models of class (i) and (ii)
are particularly popular since energetic positron due to the decay can
account for the cosmic-ray positron excess which was reported by
PAMELA~\cite{Adriani:2008zr,Adriani:2013uda},
Fermi-LAT~\cite{FermiLAT:2011ab} and
AMS-02~\cite{Aguilar:2013qda,Accardo:2014lma}, with a proper choice of
$\tau_{\rm dm}$ and $m_{\rm dm}$.  To do comprehensive analysis, we
consider several dark matter models in supersymmetric theory and
compute gamma-ray distribution for each case.

In supersymmetric models, the LSP becomes unstable once RPV operators
are introduced in the superpotential,
\begin{eqnarray}
W_{\slashed{R}_p}=\mu_i \hat{L}_i \hat{H}_u +
\lambda_{ijk}\hat{Q}_i\hat{L}_j\hat{D}^c_k
+\lambda'_{ijk}\hat{L}_i\hat{L}_j\hat{E}^c_k
+\lambda''_{ijk}\hat{U}^c_i\hat{D}^c_j\hat{D}^c_k\ ,
\end{eqnarray}
where $i,j,k$ are generation indices, $\hat{Q}_i$\,($\hat{L}_i$) is
left-handed quark\,(lepton),
$\hat{U}^c_i,\hat{D}^c_i$\,($\hat{E}^c_i$) are right-handed up- and
down-type quarks (lepton), and $\hat{H}_u$ is up-type Higgs. We use
`hat' to represent chiral superfields (superpartners are expressed by
using `tilde' in the later discussion).  The final state of dark
matter decay depends on which of the above RPV interaction terms are
operative.  For the purpose mentioned above, we consider wino,
sneutrino, gravitino and axino as the LSP. Simulated gamma-ray fluxes
are summarized in Fig.~\ref{fig:spectrum}, for which we will give the
details below.

\subsubsection{Wino}

Winos are superpartner of $W$ bosons. In a wide range of
anomaly-mediated supersymmetry breaking
scenario~\cite{Randall:1998uk}, neutral wino $\tilde{W}^0$ is the LSP
and it is a viable candidate for dark matter with a mass of a few
hundred GeV~\cite{Moroi:1999zb} or around $3~{\rm
  TeV}$~\cite{Hisano:2006nn}. Besides, scalar partners of
quarks/leptons are much heavier than TeV scale in anomaly mediation,
which is consistent with 126~GeV Higgs boson discovered at the
LHC~\cite{Aad:2012tfa,Chatrchyan:2012ufa}. Testing wino dark matter is
pointed out to be promising in direct~\cite{Hisano:2010fy} or
indirect~\cite{Hisano:2003ec} detection experiments. More recently
cosmic rays from decaying wino is studied in $LLE^c$
RPV~\cite{Ibe:2014qya} to explain the latest cosmic-ray positron
excess reported in AMS-02~\cite{Aguilar:2013qda,Accardo:2014lma}. We
study the same type of RPV considered in Ref.\,\cite{Ibe:2014qya} to
simulate gamma ray.

Since squarks are much heavier than winos, it is instructive to write
down interactions of $\tilde{W}^0$ which are relevant for the decay in
higher dimension operators. In large slepton limit, they are given by
\begin{eqnarray}
  {\cal L}_{\rm int} =
  \frac{\lambda'_{ijk}}{\Lambda^2}
  \left[
    (\bar{\tilde{W}}^0P_L e_j)(\bar{e}_kP_L\nu_j)+
    (\bar{\tilde{W}}^0P_L \nu_j)(\bar{e}_kP_Le_j)
    \right]+{\rm h.c.}
\end{eqnarray}
where $P_L=(1-\gamma_5)/2$,
$1/\Lambda^2=\sqrt{2}g_2/m_{\tilde{L}_j}^2$, and summation of flavor
index should be taken for $i<j$. $g_2$ is $SU(2)_L$ gauge coupling
constant and $m_{\tilde{L}_j}$ is soft supersymmetry breaking mass for
left-handed slepton. We have assumed $m_{\tilde{L}_j}<m_{\tilde{L}_i}$
for $i<j$. Wino decay as $\tilde{W}^0\rightarrow \nu_i e_j^-e_k^+$
($\bar{\nu}_i e_j^+e_k^-$) and $\nu_j e_i^-e_k^+$ ($\bar{\nu}_j
e_i^+e_k^-$). If lepton masses are ignored, then the energy
distribution of final state leptons is simply given by
\begin{align}
    & \frac{d^2\Gamma_{\tilde{W}^0\rightarrow \nu_i e_j^-e_k^+}}{dz_jdz_k}
    =\frac{d^2\Gamma_{\tilde{W}^0\rightarrow \bar{\nu}_i e_j^+e_k^-}}{dz_jdz_k}
    =\frac{|\lambda'_{ijk}|^2}{512\pi^3}\frac{m_{\tilde{W}^0}^5}{\Lambda^4}
    \, z_j(1-z_j), \\
    &\frac{d^2\Gamma_{\tilde{W}^0\rightarrow \nu_j e_i^-e_k^+}}{dz_idz_k}
    =\frac{d^2\Gamma_{\tilde{W}^0\rightarrow \bar{\nu}_je_i^+e_k^-}}{dz_idz_k}
    =\frac{|\lambda'_{ijk}|^2}{512\pi^3}\frac{m_{\tilde{W}^0}^5}{\Lambda^4}
    \, z_j(1-z_j),
\end{align}
where $m_{\tilde{W}^0}$ is wino mass and
$z_\alpha=2E_\alpha/m_{\tilde{W}^0}$ ($\alpha=i,j,k$) are defined by
the energy $E_\alpha$ of $ \nu_i, e_j^-,e_k^+$ and so on. They satisfy
$z_i+z_j+z_k=2$ and $0\le z_{i,j,k}\le 1$ (see also
Ref.\,\cite{Ishiwata:2009vx}). Then the energy distribution for charged
leptons in the final state is given by
\begin{align}
  \frac{dN_{e_j}}{dz_{e_j}}=12z_{e_j}^2(1-z_{e_j})\ , \ \ 
  \frac{dN_{e_k}}{dz_{e_k}}=2z_{e_k}^2(3-2z_{e_k})\ ,
  \label{eq:dN/dz_lle1}
\end{align}
 in a single process $\tilde{W}^0\rightarrow \nu_i e_j^-e_k^+$, and
\begin{eqnarray}
\frac{dN_{e_\beta}}{dz_{e_\beta}}=2z_{e_\beta}^2(3-2z_{e_\beta})\ ,
  \label{eq:dN/dz_lle2}
\end{eqnarray}
where $\beta=i,k$ in $\tilde{W}^0\rightarrow \nu_j e_i^-e_k^+$. Energy
distribution for each particle in CP conjugated final state is the
same as in original state. For example, if $\lambda_{122}$ is only
relevant, then the final state in wino decay is $\nu_e \mu^-\mu^+$
($\bar{\nu}_e \mu^+\mu^-$) and $\nu_\mu e^-\mu^+$ ($\bar{\nu}_\mu
e^+\mu^-$). Thus no primary gamma rays or (anti-)protons are
generated. In such case, IC photons from $e^{\pm}$ (which includes
electron and positron from $\mu^\pm$) are the only observable gamma
rays. If $\tau^\pm$ is produced, then its cascade decay produce
energetic gamma rays.  We use PYTHIA~\cite{Sjostrand:2006za} to
compute $dN_I/dE$ ($I=\gamma,\,e^{\pm}$) in the primary decay of dark
matter.

\subsubsection{Sneutrino}

As a reference, we also consider right-handed sneutrino dark matter in
$LLE^c$ RPV.  Supposing that neutrinos are purely-Dirac fermions,
neutrino Yukawa couplings are very small, i.e.  ${\cal O}(10^{-13})$.
Since right-handed sneutrinos, superpartners of right-handed
neutrinos, interact with the other particles via the Yukawa couplings,
the lightest one can be dark matter if it is the
LSP~\cite{Asaka:2005cn}.  In $LLE^c$ type RPV, right-handed sneutrino
$\tilde{\nu}_{Ri}$ decays to charged leptons $l^+_jl^-_k$. We will
compute gamma-ray flux for the final state
$l^+_jl^-_k$.\footnote{Electroweak correction to the final state
  becomes important when dark matter mass is extremely large. However,
  we have checked numerically that it does not affect the result when
  the mass is less than $10\,{\rm TeV}$.}

\subsubsection{Gravitino}

In $R$-parity conserved case, gravitino LSP is typically disfavored in
cosmology when it has electroweak-scale mass. This is because the
next-LSP (NLSP), which is usually the LSP in the minimal
supersymmetric standard model and decays to gravitino and
standard-model particles via Planck-suppressed interaction, becomes
long-lived and may decay after big-bang nucleosynthesis (BBN)
started. In the RPV case, however, NLSP can decay to standard-model
particles via RPV interaction before BBN starts. Thus NLSP decay does
not affect BBN. Furthermore, gravitino is cosmologically long-lived
since the decay rate is suppressed by the Planck mass and a small
violation of $R$-parity, thus it can play the role of dark
matter~\cite{Takayama:2000uz}.  In bi-linear RPV, for example,
gravitino $\psi_{3/2}$ decays to $\gamma\nu_i$, $W^{\pm}l_i^{\mp}$,
$Z\nu_i$ and $h\nu_i$ if kinematically allowed. It is shown that the
branching fraction of $\psi_{3/2}\rightarrow W^{\pm}l_i^{\mp}$ is the
largest~\cite{Ishiwata:2008cu}, which means that the decay products
after cascading decay of $ W^{\pm}l_i^{\mp}$ are mixture of
high-energy electrons/positrons, gamma rays and
protons/anti-protons.\footnote{Neutrinos are also produced, but they
  are irrelevant in our discussion.} Recently decaying gravitino dark
matter is revised under bi-linear
RPV~\cite{Ibe:2013nka,Carquin:2015uma} to account for cosmic-ray
positron excess observed in AMS-02 experiments.  In our analysis we
simply consider $\psi_{3/2}\rightarrow W^{\pm}l_i^{\mp}$ to avoid
introducing parameters involved in determination of each branching
fraction, and simulate primary and IC gamma rays. Inclusion of the
other channels is straightforward (see Ref.\,\cite{Ishiwata:2008cu}),
and that will not change our result significantly.

\subsubsection{Axino}

Axino, a superpartner of axion originally introduced to solve the
strong CP problem~\cite{Peccei:1977hh}, can be dark matter in some
supersymmetry models. In addition to that, if baryon number is broken
by $U^cD^cD^c$ term with $\lambda''_{ijk} \lesssim 1$,\footnote{See
  Refs.\,\cite{Barbier:2004ez,Bhattacherjee:2013gr} for realization of
  such case.} baryon asymmetry is generated by moduli decay before BBN
to give the observed baryon number~\cite{Ishiwata:2013waa}. Even in
such a large RPV, it is shown that axino with a mass of ${\cal
  O}(10\,{\rm GeV})$ can be long-lived to be dark
matter~\cite{Ishiwata:2014cra}.

In this model, scalar superpartners are much heavier than the
electroweak scale. The interactions of axino (denoted as $\tilde{a}$)
with squarks are then written in higher dimension operators. Ignoring
down-type squarks and left-right mixing in squark sector for
simplicity, it is given by
\begin{eqnarray} 
   {\cal L}_{\rm int}=\frac{\lambda''_{ijk}}{\Lambda^2} (\bar{u}_i P_L
   \tilde{a})(\bar{d}_kP_Ld_j^c) + \,{\rm h.c.}\ ,
\end{eqnarray}
where $1/\Lambda^2 =2g_{\rm eff}^R/m_{\tilde{u}_{R_i}}^2$. Here
$m_{\tilde{u}_{R_i}}$ is supersymmetry breaking mass of
$\tilde{u}_{R_i}$, and $g_{\rm eff}^R$ is one of the dimensionless
couplings of axino-quark-squark interaction, which is suppressed by
axino decay constant (see Refs.\,\cite{Covi:1999ty,Chun:2011zd}). Color
indices are implicit. Then axino decays in three-body process:
$\tilde{a} \rightarrow u_id_jd_k$, $\bar{u}_i\bar{d}_j\bar{d}_k$. If
quark masses are ignored, then the energy distribution of final state
quarks is simply given by
\begin{eqnarray}
    \frac{d^2\Gamma_{\tilde{a} \rightarrow u_id_jd_k}}{dz_idz_j}
    =\frac{d^2\Gamma_{\tilde{a} \rightarrow
        \bar{u}_i\bar{d}_j\bar{d}_k}}{dz_idz_j}
    =\frac{3|\lambda''_{ijk}|^2}{128\pi^3}\frac{m_{\tilde{a}}^5}{\Lambda^4}
    \, z_i(1-z_i),
\end{eqnarray}
where $m_{\tilde{a}}$ is axino mass, $z_i=2E_i/m_{\tilde{a}}$ and
$z_j=2E_j/m_{\tilde{a}}$ are defined by the energy $E_i$ and $E_j$ of
$u_i$ and $d_j$, respectively.  Then it is straightforward to give the
energy distribution of each final sate quark:
\begin{align}
  \frac{dN_i}{dz_i}=12z_i^2(1-z_i)\ , \ \ 
  \frac{dN_j}{dz_j}=2z_j^2(3-2z_j)\ ,
  \label{eq:dN/dz_udd}
\end{align}
in a single process $\tilde{a}\rightarrow u_id_jd_k$.  The energy
distribution for $d_k$ is the same as $d_j$. These quarks are
hadronized to produce mesons, which decay to gamma rays and
electrons/positrons, and electrons/positrons become source of IC
photons. In later numerical analysis, we also compute a case of final
state $b\bar{b}$ for comparison, which would be useful for those who
are interested in.

\subsection{Gamma-ray fluxes (examples)}

\begin{figure}
 \begin{center}
  \includegraphics[width=10cm]{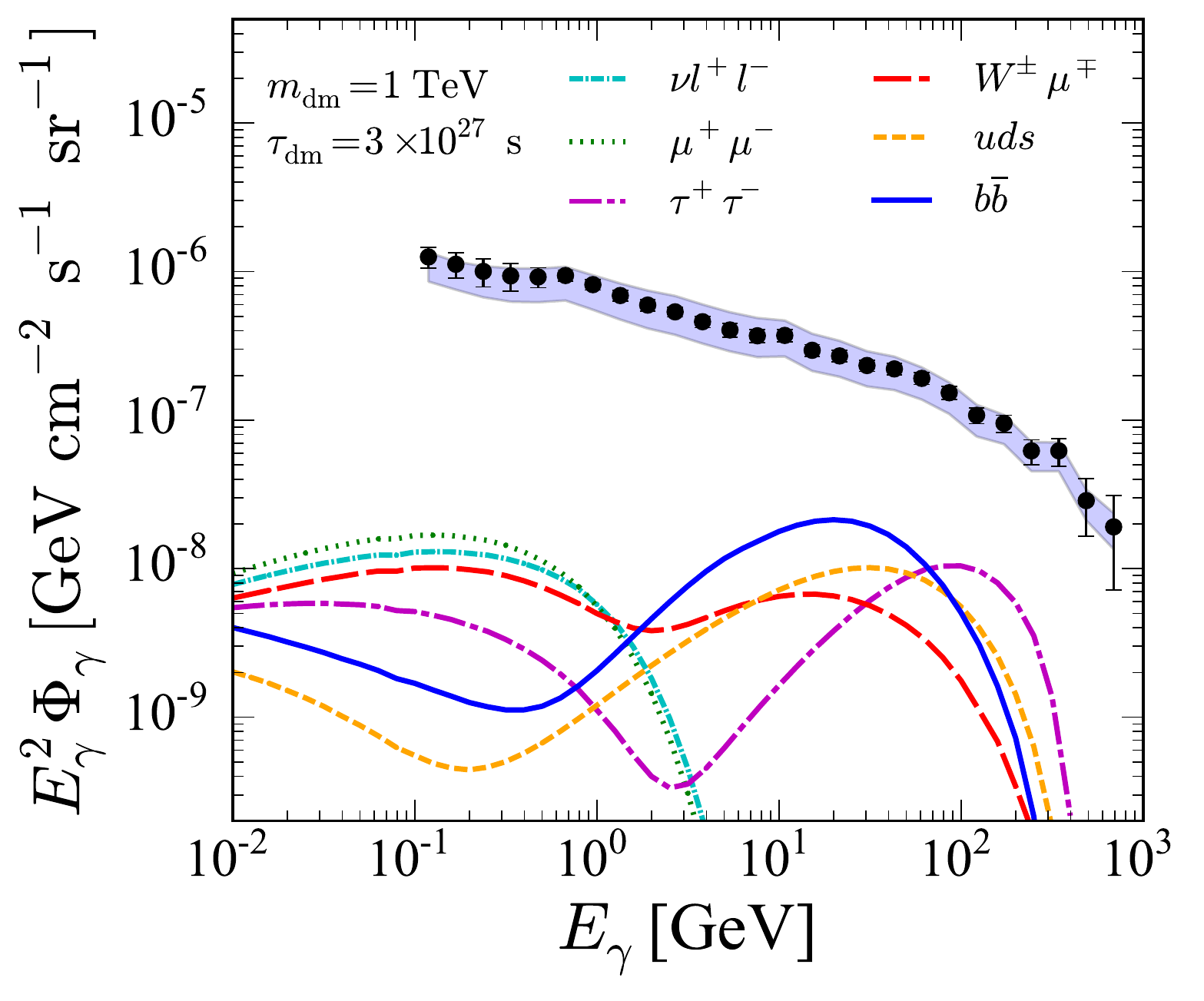}
  \caption{Gamma-ray fluxes from various decaying dark matter ($m_{\rm
      dm} = 1$~TeV, $\tau_{\rm dm} = 3 \times 10^{27}$~s). Plots give
    fluxes from decay channels: (a) $\nu_e \mu^-\mu^+$ ($\bar{\nu}_e
    \mu^+\mu^-$) and $\nu_\mu e^-\mu^+$ ($\bar{\nu}_\mu e^+\mu^-$),
    (b) $\mu^+\mu^-$, (c) $\tau^+\tau^-$, (d) $W^{\pm}\mu^{\mp}$, (e)
    $uds$ ($\bar{u}\bar{d}\bar{s}$), and (f) $b\bar{b}$.  Data points
    with error bar and a band of the EGRB observed by Fermi-LAT is
    also shown~\cite{Ackermann:2014usa} (see Sec.~\ref{sec:astro}).}
  \label{fig:spectrum}
 \end{center}
\end{figure}

In Fig.~\ref{fig:spectrum} gamma-ray fluxes in various decaying dark
matter models are plotted. For leptophilic case, result is shown for a
case where only $\lambda'_{122}$ is relevant (dubbed as ``$\nu
l^+l^-$'') in $\tilde{W}^0$ dark matter, while decay channels
$\mu^+\mu^-$ and $\tau^+\tau^-$ are considered in $\tilde{\nu}_R$
decay. It is seen that the gamma-ray spectra from $LLE^c$ and
$\mu^+\mu^-$ are quite similar. On the other hand, in $\tau^+\tau^-$,
the spectrum has double peaks. This is due to primary gamma rays
produced from cascade decay of tau, which gives another gamma-ray flux
in high energy region. For hadronically decaying dark matter, the
axino decay via $\lambda''_{122}$ is considered (denoted as
``$uds$''). The spectrum shows similar behavior to $\tau^+\tau^-$ case
and $b\bar{b}$ channel as well. Finally, the flux from decaying
gravitino to $W^{\pm}\mu^{\mp}$ is expected to have a property in the
middle of leptophilic and hadrophilic cases, which is in fact seen in
the figure.

\section{Astrophysical source}
\label{sec:astro}
\setcounter{equation}{0} 

Conventional astrophysical sources emit gamma rays through particle
acceleration followed by interactions between accelerated charged
particles and surrounding media or photon fields.
The total EGRB should be made of both dark matter and astrophysical
components:
\begin{equation}
\Phi_\gamma(E_\gamma) = \Phi_\gamma^{\rm astro}(E_\gamma) +
 \Phi_\gamma^{\rm dm} (E_\gamma)\ .
\end{equation}
In this section we discuss three possible astrophysical sources in
extragalactic region.  They are blazars, star-forming galaxies (SFGs),
and misaligned active galactic nuclei (mAGNs).

\subsection{Blazars}

In the GeV gamma-ray sky, the most dominant source is blazars, which
are one class of active galaxies whose jets are directed towards us.
More than thousand blazars have been identified with
Fermi-LAT~\cite{FermiCatalog} and they indeed make up a significant
fraction ($\sim$70\%) of the total measured EGRB
intensity~\cite{Ackermann:2014usa, Ajello:2015mfa}.  Therefore, it is
essentially important to include this source class with a particular
care.

\begin{figure}
 \begin{center}
  \includegraphics[width=10cm]{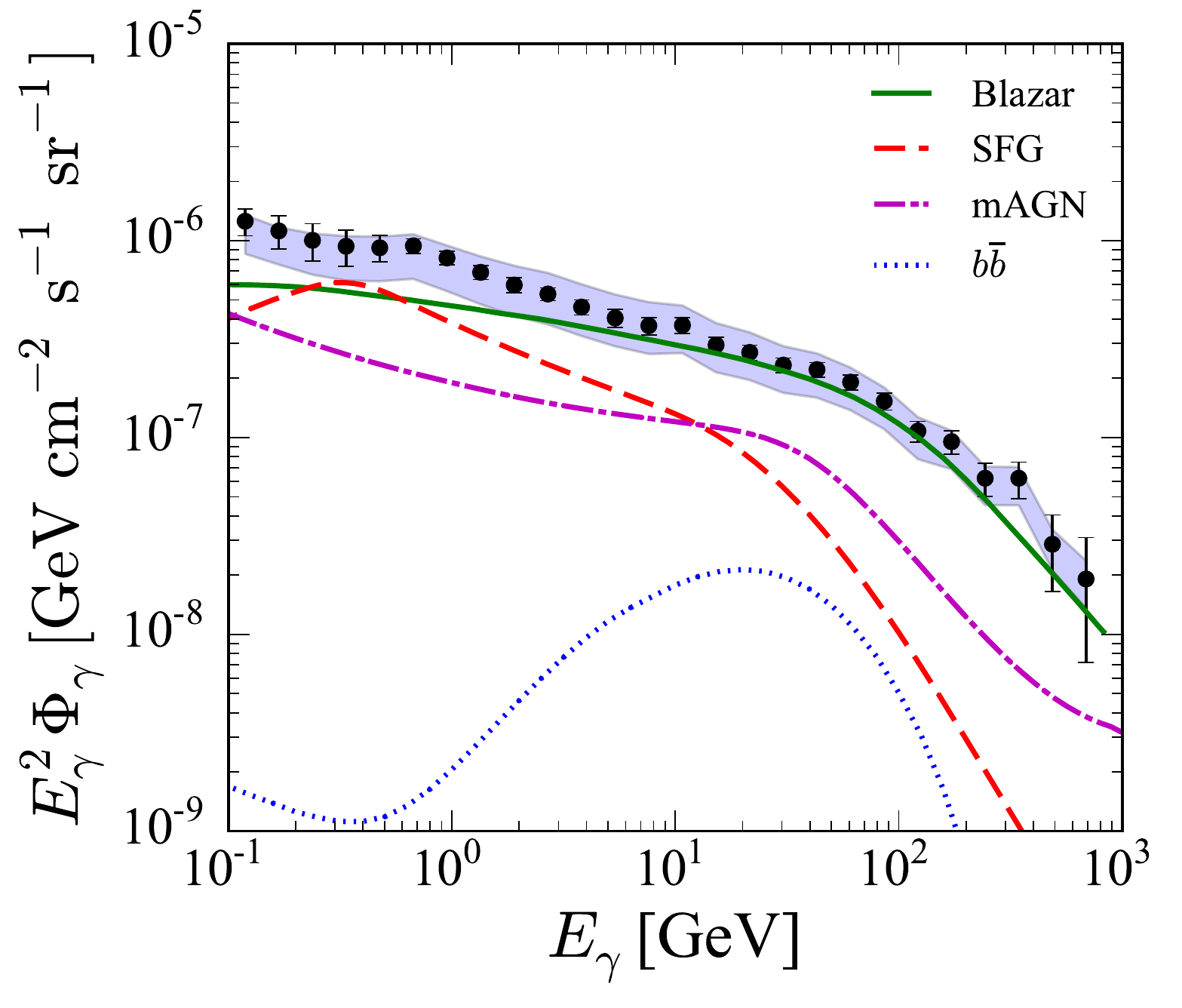}
  \caption{EGRB spectrum from blazars (solid), star-forming galaxies
    (dashed), misaligned active galaxies (dot-dashed), and dark matter
    decay (dotted; $m_{\rm dm} = 1$~TeV, $\tau_{\rm dm} = 3 \times
    10^{27}$~s, $b\bar{b}$ decay channel).  Data with error bars show
    the latest Fermi-LAT measurement~\cite{Ackermann:2014usa}, and a
    band attached to them show systematic uncertainty due to
    foreground subtraction (see
    Sec.~\ref{sec:ConstraintsOnTauDM}). Note these data are {\it
      total} EGRB, which includes already resolved sources. {\it
      Unresolved} EGRB (or also referred to as {\it isotropic}
    gamma-ray background), on the other hand, is about a factor of two
    smaller~\cite{Ackermann:2014usa}.}
  \label{fig:spectrum_bb}
 \end{center}
\end{figure}

Since many blazars have been identified with known redshifts, it is
possible to reconstruct its gamma-ray luminosity function (i.e.,
number density of blazars per unit luminosity range).  This has been
done extensively, and thus far, the luminosity-dependent density
evolution model that is motivated by possible underlying correlations
between X-ray and gamma-ray luminosities gives the most reasonable fit
to both the luminosity and redshift distributions.  The luminosity
functions constructed this way can be found in Refs.\,\cite{FSRQ,
  BLLac} for flat-spectrum radio quasars and BL Lactae objects,
respectively (each is a subclass of the blazar population).  In this
study, we adopt the EGRB spectrum due to blazar population from
Ref.\,\cite{Ajello:2015mfa} based on the latest reconstruction of the
blazar luminosity function, which is shown in
Fig.~\ref{fig:spectrum_bb}.  The current uncertainty range of the
spectrum is about $\sim$30\%, almost independent of energy.
Fig.~\ref{fig:spectrum_bb} shows that the blazars makes up of
significant fraction of the EGRB.  This is not very surprising because
the EGRB data shown in the figure also includes all the resolved
extragalactic sources, most of which are blazars.  However, it is
quite clear that they cannot be the {\it only} source that can explain
the EGRB in the entire energy region between 100~MeV and 820~GeV.

\subsection{Star-forming galaxies}

SFGs are galaxies that are undergoing active star formation just like
our own Milky-Way Galaxy.  Even though they are in general much dimmer
than the blazars, already several of them are detected with
Fermi-LAT~\cite{SFG}.  The same study also found a tight correlation
between the gamma-ray luminosity and infrared luminosity for these
identified galaxies.  Assuming that this relation universally holds
and by using the precise infrared luminosity function of SFGs, one is
able to construct the gamma-ray luminosity function of SFGs.
Reference~\cite{Tamborra} computed the EGRB intensity due to SFGs,
starburst galaxies, and SFGs containing AGNs separately, by adopting
the latest infrared luminosity function constructed with the Herschel
telescope~\cite{Herschel}.  The same paper found that the uncertainty
of the EGRB spectrum came mainly from those due to gamma-ray--infrared
luminosity correlation, and it yielded about $\sim$60\% uncertainties.
It also yielded larger flux than the previous model~\cite{SFG}.
Fig.~\ref{fig:spectrum_bb} shows the SFG contribution (that also
includes contributions from starbursts and SFGs containing AGNs).  The
peak around $\sim$0.3--0.5~GeV comes from kinematical argument of the
pion decay that is produced in the interactions between cosmic rays
and interstellar medium.  The spectrum becomes suppressed above
$\sim$20~GeV because of the effect of gamma-ray absorption due to the
extragalactic background light.  The SFG spectrum nicely complements
that of blazars especially at low energies.

\subsection{Misaligned active galactic nuclei}

Yet another population, although much less certain, is the mAGNs,
whose jets are not directed towards the Earth.  (They are also known
as radio galaxies.)  Even though they are less bright than the blazars
in gamma rays, a few of them are identified with Fermi-LAT.  The
gamma-ray luminosity function for this population is constructed
with the radio luminosity function and measured correlation between
radio and gamma-ray luminosities.  Although with much larger
uncertainties, this has been done in the literature~\cite{mAGN1,
  mAGN2}, and for this work, we adopt the model of Ref.\,\cite{mAGN1}
that is also shown in Fig.~\ref{fig:spectrum_bb}.  The uncertainty is
still quite large due to paucity of the detected sources, which is
typically a factor of three.

\section{Results}
\label{sec:result}
\setcounter{equation}{0} 

We give the lower bound for the lifetime of dark matter.  Based on the
discussion in Sec.~\ref{sec:dm}, we consider following final state in
dark matter decay; $\nu_e \mu^-\mu^+$ ($\bar{\nu}_e
\mu^+\mu^-$) \& $\nu_\mu e^-\mu^+$ ($\bar{\nu}_\mu e^+\mu^-$),
$\mu^+\mu^-$, $\tau^+\tau^-$, $W^{\pm}\mu^{\mp}$, $uds$
($\bar{u}\bar{d}\bar{s}$), $b\bar{b}$.

\subsection{Analysis}

With the EGRB data measured with Fermi-LAT~\cite{Ackermann:2014usa} and
both the dark matter and astrophysical models discussed in the previous
sections, we perform the Bayesian statistical analysis using the
Markov-Chain Monte Carlo simulation to obtain 95\% credible lower limit
on dark matter decay lifetime for given dark matter model characterized
by mass and decay channel.

The posterior probability distribution of parameters
${\bm\vartheta}$ given data ${\bm d}$, $P({\bm\vartheta} | {\bm d})$, is
related to the prior $P({\bm\vartheta})$ and the likelihood function
$\mathcal{L}({\bm d} | {\bm \vartheta})$ through
\begin{equation}
 P({\bm \vartheta} | {\bm d}) \propto P({\bm \vartheta})
  \mathcal{L}({\bm d} | {\bm \vartheta})\ .
\end{equation}
We assume that the likelihood function $\mathcal L({\bm d} | {\bm
\vartheta})$ is approximated to be normal distribution
\begin{equation}
\mathcal L({\bm d} | {\bm\vartheta}) = \prod_i
 \frac{1}{\sqrt{2\pi}\sigma_{\Phi_i}}
 \exp\left(-\frac{[\Phi_i - \Phi_\gamma^{\rm dm}(E_i | \tau_{\rm dm}) -
      \Phi_\gamma^{\rm
      astro}(E_i | f_{\rm blazar}, f_{\rm SFG}, f_{\rm mAGN})]^2}{2
      \sigma_{\Phi_i}^2}\right)\ ,
\label{eq:likelihood}
\end{equation}
where $E_i$ and $\Phi_i$ represent the energy and the EGRB intensity
data in $i$-th bin, respectively.
The $1\sigma$ errors for $\Phi_i$, assumed to be uncorrelated between
different energy bins, are shown as $\sigma_{\Phi_i}$ (and also as the
error bars in Fig.~\ref{fig:spectrum_bb}).
For theoretical parameters, we adopt the dark matter lifetime, $\tau_{\rm
dm}$, and energy-independent normalization for each of the astrophysical
components, $f_{\rm blazar}$, $f_{\rm SFG}$, and $f_{\rm mAGN}$, for
which $f = 1$ recovers our reference spectra shown in
Fig.~\ref{fig:spectrum_bb}.

For the prior $P(\tau_{\rm dm})$, we adopt a flat distribution in
$\log\tau_{\rm dm}$ in the range of $20 < \log (\tau_{\rm dm} / {\rm
  s}) < 30$, which covers quite a wide range without any bias.
Regarding the prior for the astrophysical parameters, we adopt normal
distributions centered on $\log f = 0$, and with standard deviation of
$\sigma_{\log f_{\rm blazar}} = 0.1$, $\sigma_{\log f_{\rm SFG}} =
0.2$, and $\sigma_{\log f_{\rm mAGN}} = 0.5$, based on the errors
$25\%$, $60\%$, and $200\%$, for the blazars, SFGs, and mAGNs,
respectively. Although the errors of blazar depend on the energies of
gamma ray, they are $\sim$25\% in most energy bins, while they are
larger ($\sim$30\%) at a few high-energy bins.  These `Normal' priors,
also summarized Table~\ref{table:astro prior}, accommodate the best
knowledge thus far obtained for the astrophysical sources as discussed
in Sec.~\ref{sec:astro}.  We note that since the information of the
resolved sources are already used in order to obtain these Normal
priors, here we adopt the errors on {\it unresolved} EGRB
intensity~\cite{Ackermann:2014usa} in denominator of
Eq.~(\ref{eq:likelihood}). For the gamma-ray intensity, on the other
hand, we use the data of the total EGRB.

\begin{table}
  \begin{center}
   \caption{Priors for dark matter and astrophysical parameters.}
   \label{table:astro prior}
  \begin{tabular}{|ccccc|}\hline
   Prior & $\tau_{\rm dm}$ & $f_{\rm blazar}$ & $f_{\rm SFG}$ & $f_{\rm
   mAGN}$ \\ \hline
   Normal & $20 < \log (\tau_{\rm dm}/{\rm s})< 30$ & $\sigma_{\log f} =
	   0.1$ & $\sigma_{\log f} = 0.2$ & $\sigma_{\log f} = 0.5$ \\
   Flat & $20 < \log (\tau_{\rm dm}/{\rm s})< 30$ & $-5 < \log f < 2$ &
	       $-5 < \log f < 2$ &  $-5 < \log f < 2$ \\ \hline
  \end{tabular}
  \end{center}
\end{table}

In order to give more conservative limits, we also discuss the case
without any astrophysical prior information, for which we simply adopt
the log-flat priors for all $f$ parameters, between $-5 < \log f < 2$.
These `Flat' priors are summarized in Table~\ref{table:astro prior}.
Since $f$ parameters follow the log-flat distribution down to a very
small value ($f = 10^{-5}$), the parameter space includes a case where
the astrophysical sources give virtually zero contribution. Thus, in
order to analyze consistently, we adopt the data $\Phi_i$ and errors
$\sigma_{\Phi_i}$ for the unresolved EGRB (see caption of
Fig.~\ref{fig:spectrum_bb}).  We note that this approach is somewhat
pessimistic, because many astrophysical sources are known to give
contributions to the unresolved part of the EGRB, and hence taking
extremely small values for $f$ will result in significant
underestimate of these astrophysical components.

For purely phenomenological purpose and also in comparison to the
results with 10-month Fermi data in the
literature~\cite{Cirelli:2012ut}, we also adopt a single cutoff
power-law component for non dark matter EGRB.  Although from
Fig.~\ref{fig:spectrum_bb} such a single cutoff power-law model
appears to give reasonable fit to the Fermi data (indeed it
does~\cite{Ackermann:2014usa}), it is no longer strongly motivated
since the resolved components (mainly blazars) are already shown to
feature quite different spectral shape than the EGRB data.  For this
phenomenological power-law model (labeled as `PL'), we adopt the
normalization factor $f_{\rm PL}$, where $f_{\rm PL} = 1$ corresponds
to the best fit parameter to the data on its own, the spectral index
$\gamma$, and the exponential cutoff energy $E_{\rm cut}$, following
Ref.\,\cite{Ackermann:2014usa}.  For $f_{\rm PL}$, we use the log-flat
prior, and for the other two flat priors in linear scale.  The ranges
of these priors are summarized in Table~\ref{table:PL prior}.

\begin{table}
   \begin{center}
    \caption{Priors for phenomenological parameters for `astrophysical'
    component.}
    \label{table:PL prior}
  \begin{tabular}{|cccc|}\hline
   Prior & $f_{\rm PL}$ & $\gamma$  & $E_{\rm cut}$ \\ \hline
   PL & $-2 < \log f < 2$ & $-3.5 < \gamma < -1$ & $100 < E_{\rm cut} /
	       {\rm GeV} < 600$ \\ \hline
  \end{tabular}
   \end{center}
\end{table}

\subsection{Constraints on dark matter lifetime}
\label{sec:ConstraintsOnTauDM}
\begin{figure}
 \begin{center}
   \includegraphics[width=15cm]{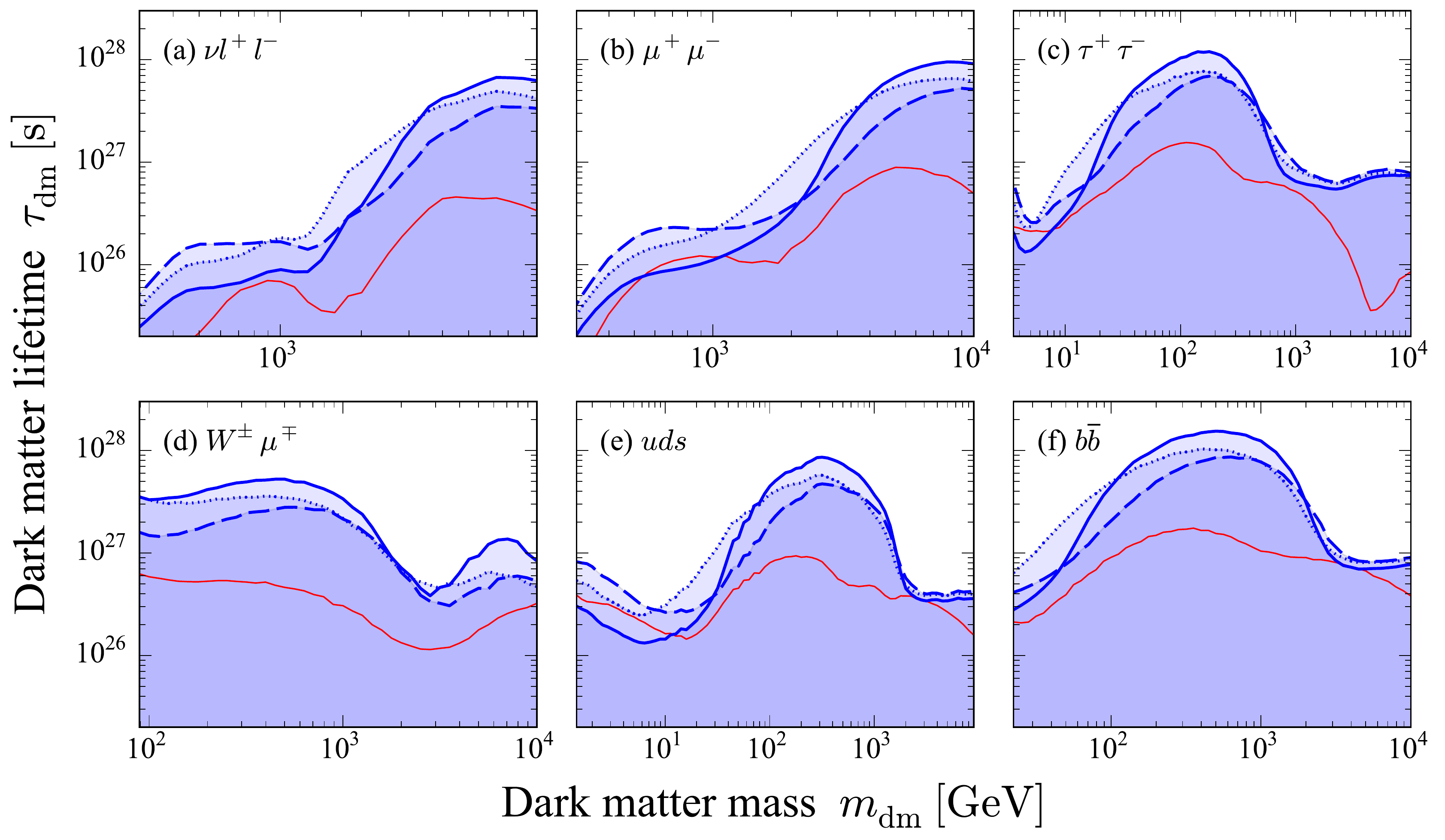}
  \caption{95\% credible lower limits on dark matter lifetime
    $\tau_{\rm dm}$ as function of dark matter mass $m_{\rm dm}$, for
    decay channels: (a) $\nu_e \mu^-\mu^+$ ($\bar{\nu}_e \mu^+\mu^-$)
    and $\nu_\mu e^-\mu^+$ ($\bar{\nu}_\mu e^+\mu^-$), (b) $\mu^+
    \mu^-$, (c) $\tau^+\tau^-$, (d) $W^\pm\mu^\mp$, (e) $uds$
    ($\bar{u}\bar{d}\bar{s}$), (f) $b\bar{b}$. Astrophysical
    background models with Normal priors are adopted
    (Table~\ref{table:astro prior}). Thick solid, dashed, and dotted
    curves correspond to the EGRB data with different foreground
    modeling discussed in Ref.\,\cite{Ackermann:2014usa} (their models
    A, B, and C, respectively). Thin solid curve shows the lower
    limits obtained with the 10-month Fermi-LAT
    data~\cite{Abdo:2010nz} and the phenomenological power-law
    background modeling.}
  \label{fig:limit_tau_Normal}
 \end{center}
\end{figure}

Figure~\ref{fig:limit_tau_Normal} shows 95\% credible lower limits on
the decay lifetime as function of its mass for various decay channels.
Here we adopted the astrophysical background models with Normal prior
(Table~\ref{table:astro prior}), which should be regarded as our
canonical results based on our most up-to-date knowledge on
astrophysical sources.  Thick solid curves in
Fig.~\ref{fig:limit_tau_Normal} shows the lifetime constraints
obtained with the data shown in Fig.~\ref{fig:spectrum_bb}.  These
data are obtained after the Galactic foreground emission being
subtracted, after it was modeled with a parameter set of the Galactic
cosmic ray propagation (model A in Ref.\,\cite{Ackermann:2014usa}.)
We find that the new lifetime constraints are much more stringent than
those obtained with the 10-month data by up to more than one order of
magnitude.

In leptonically decaying dark matter models, such as $\mu^+\mu^-$ and
$\nu_e \mu^-\mu^+$ ($\bar{\nu}_e \mu^+\mu^-$) \& $\nu_\mu e^-\mu^+$
($\bar{\nu}_\mu e^+\mu^-$), constraint on the lifetime becomes
stringent almost monotonically as $m_{\rm dm}$ and $\tau_{\rm
  dm}<10^{27}$--$10^{28}~{\rm s}$ around $m_{\rm dm}\sim 10~{\rm TeV}$
is excluded. This can be understood since gamma rays in IC process is
the only contribution from dark matter and the peak of the intensity
(i.e. $E^2_\gamma \Phi_\gamma^{\rm dm}$) shifts to the region where
Fermi-LAT data have better accuracy as $m_{\rm dm}$ increases. (See
Fig.~\ref{fig:spectrum}.) In the other cases, where decay products
contain lots of hadrons, the lifetime shorter than $10^{28}$~s is
almost excluded for dark matter masses between a few 100~GeV and
$\sim$1~TeV. As opposed to the leptophilic case, the constraints get
weaker for $m_{\rm dm}\gtrsim 1~{\rm TeV}$. This is because the
gamma-ray spectrum have two peaks which comes from the cascade decay
and one from IC process. When $m_{\rm dm}\gtrsim {\rm TeV}$, a
``valley'' in the spectrum enters in the range $1~{\rm GeV}\lesssim
E_\gamma \lesssim 100~{\rm GeV}$, where Fermi-LAT data is given with
small errors. Thus the constraint gets relaxed.\footnote{Those
  interpretation is qualitative. Quantitative behavior is determined
  by non-trivial interplay between modeling of the foreground
  subtraction and choices of the prior. See discussion below.}

Quantitative arguments depend on the foreground model adopted and
subtracted from the total gamma-ray emission.  To this end, we
repeated the same computation by using two different foreground
models, B and C adopted also in Ref.\,\cite{Ackermann:2014usa}.
Models A--C nicely covers regions shown as uncertainty band in
Fig.~\ref{fig:spectrum_bb}.  The dashed and dotted curves are the
results corresponding to models B and C, respectively.  This shows
that the foreground modelings give uncertainty on lifetime constraints
by about a factor of a few.

\begin{figure}
 \begin{center}
  \includegraphics[width=15cm]{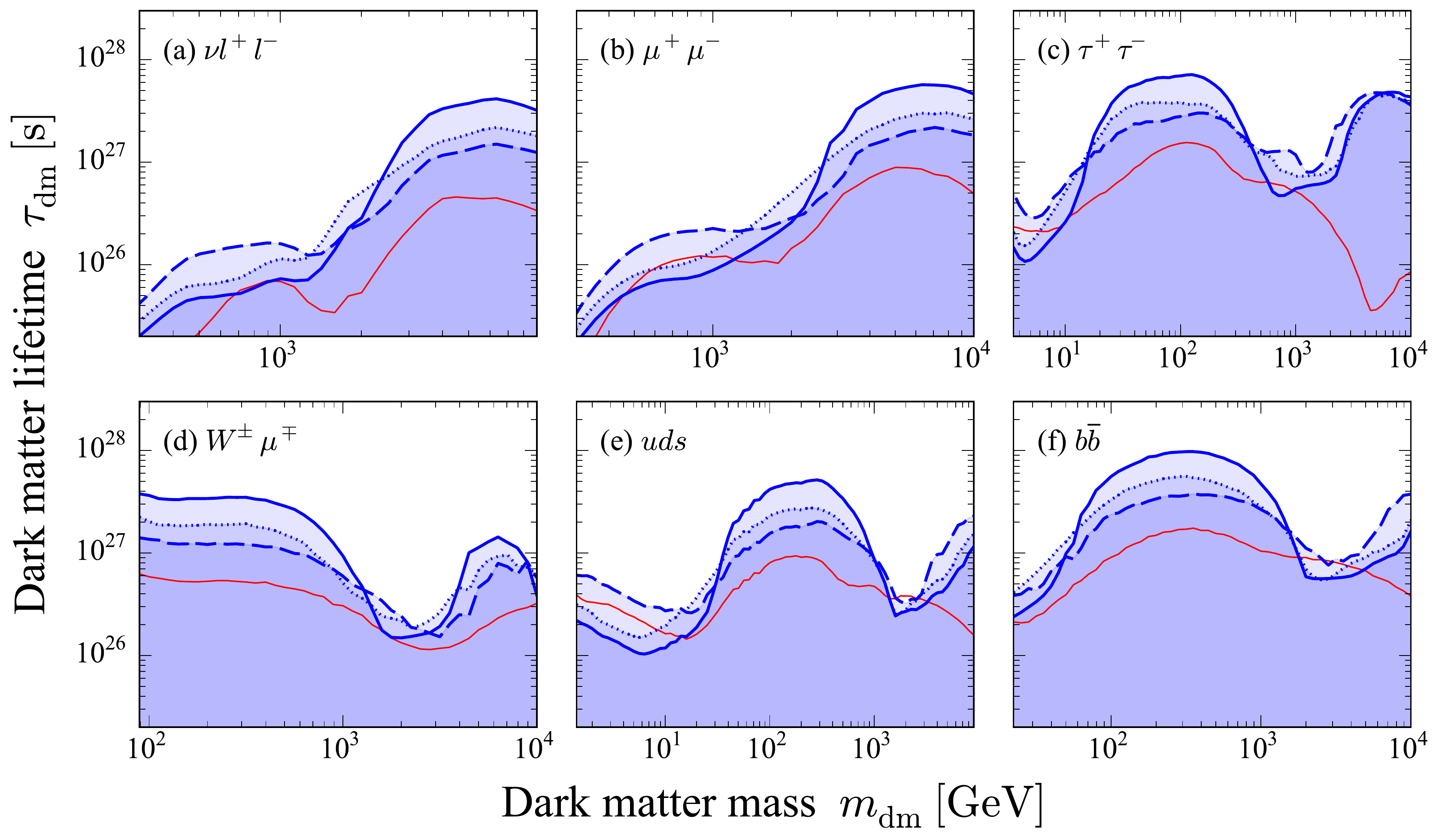}
  \caption{The same as Fig.~\ref{fig:limit_tau_Normal}, but for the
    astrophysical background models with Flat priors
    (Table~\ref{table:astro prior}). Note that these priors are very
    conservative.}
  \label{fig:limit_tau_Flat}
 \end{center}
\end{figure}

The results of more conservative approach with Flat priors in
Table~\ref{table:astro prior} are shown in
Fig.~\ref{fig:limit_tau_Flat}.  As expected, in most cases, they are
weaker than the ones with Normal priors (as shown in
Fig.~\ref{fig:limit_tau_Normal}) by about a factor of a few.
Exceptions are at high dark matter masses for (c)--(f), where they
give stronger constraints; this is likely caused by interplay between
different choices of priors and the data (the total EGRB data for the
Normal priors, while the unresolved EGRB data for the Flat priors).

\begin{figure}
 \begin{center}
  \includegraphics[width=15cm]{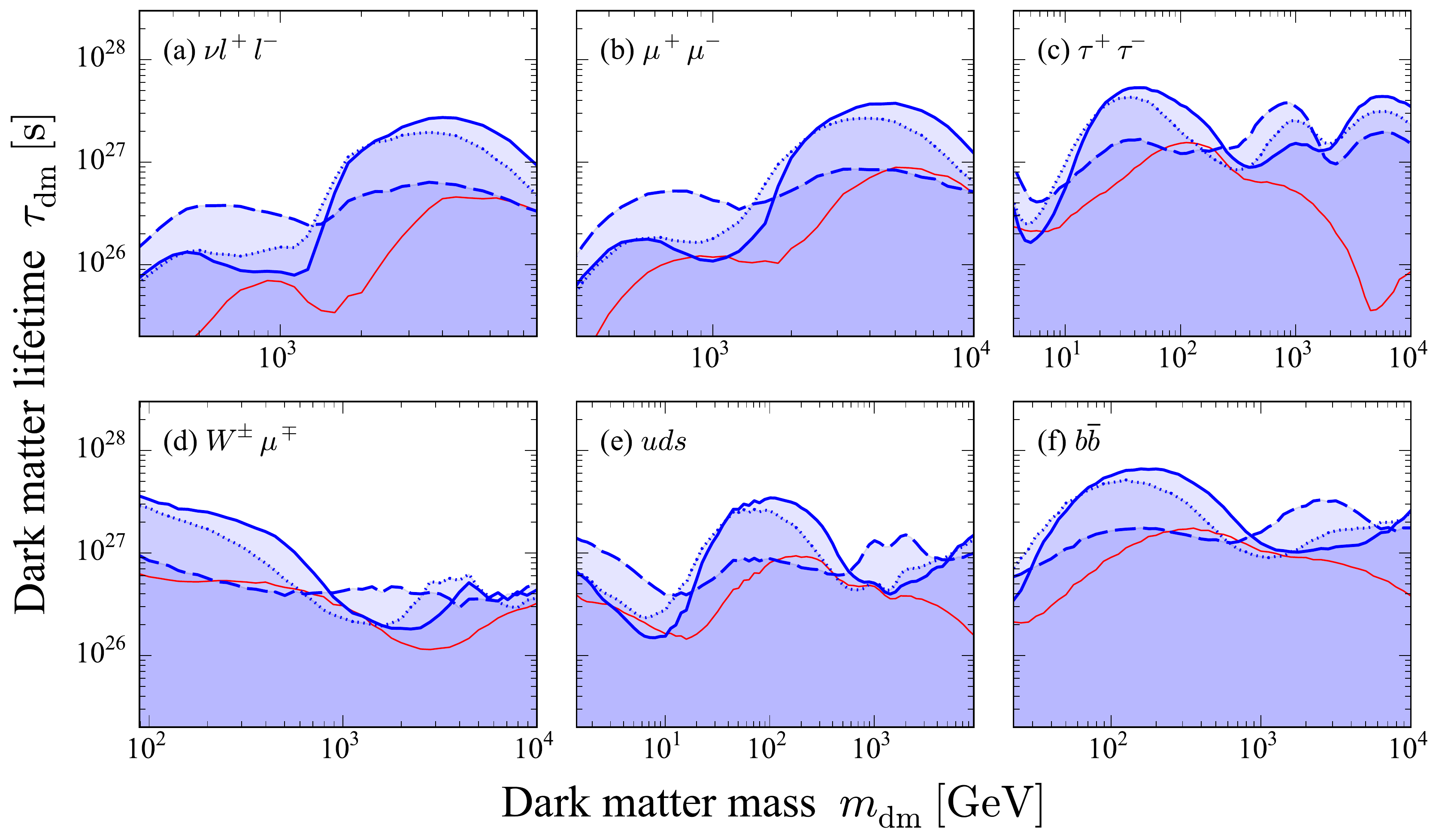}
  \caption{The same as Fig.~\ref{fig:limit_tau_Normal}, but for
  phenomenological background models with PL prior (Table~\ref{table:PL
  prior}).}
  \label{fig:limit_tau_PL}
 \end{center}
\end{figure}

\begin{figure}
 \begin{center}
   \includegraphics[width=15cm]{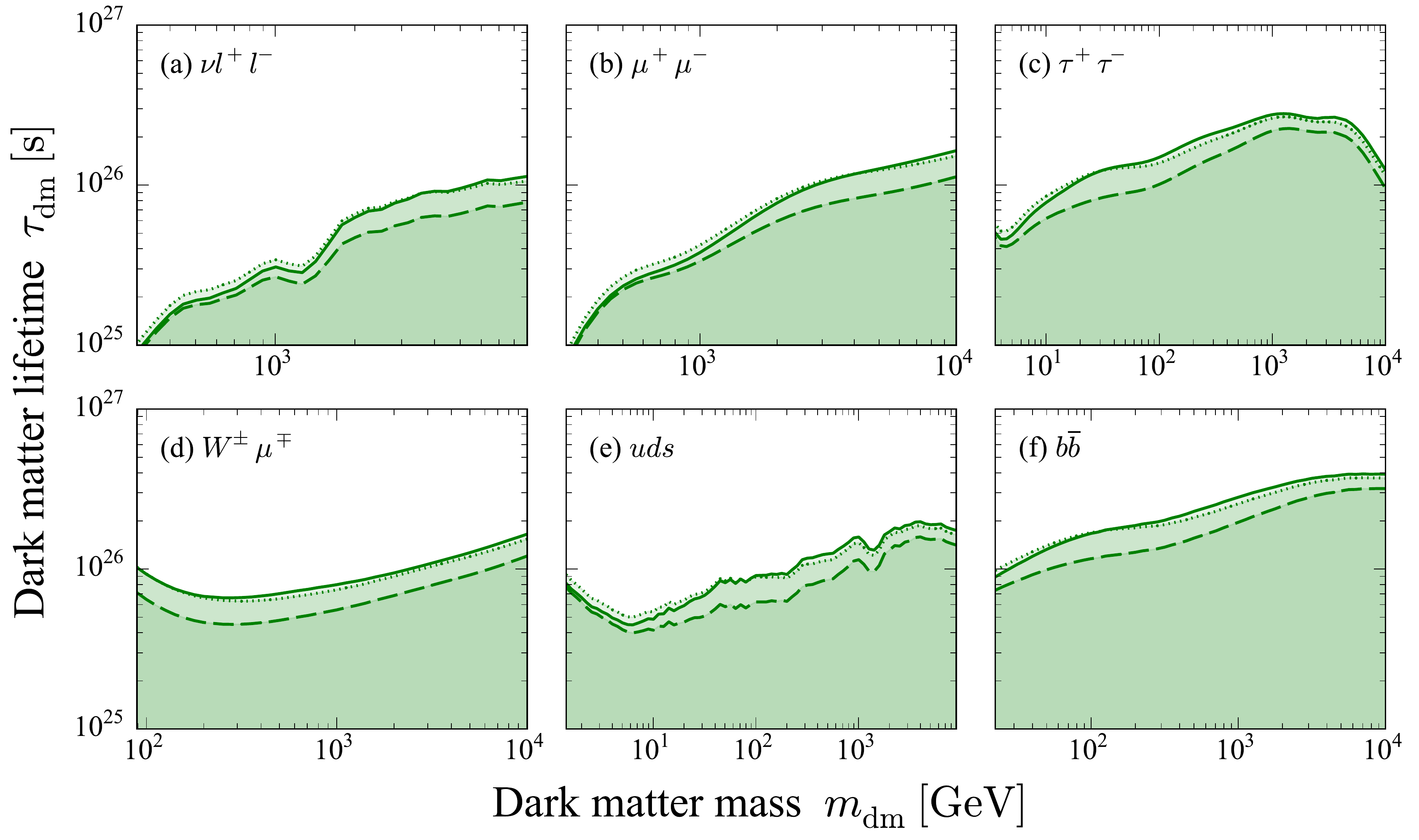}
   \caption{ The same as Fig.~\ref{fig:limit_tau_Normal}, but taking
     into account only dark matter contribution.}
  \label{fig:limit_DMonly}
 \end{center}
\end{figure}

In order to compare our results with the previous ones in the
literature (e.g., Ref.\,\cite{Cirelli:2012ut}), we also computed the
lifetime constraints by using the 10-month Fermi-LAT
data~\cite{Abdo:2010nz}.  Here we modeled the other background
component as a single power law (Table~\ref{table:PL prior}), and the
results are shown as a thin curve in each panel of
Figs.~\ref{fig:limit_tau_Normal},\ref{fig:limit_tau_Flat} and
\ref{fig:limit_tau_PL} for reference.  Although the statistics adopted
here is different than that in Ref.\,\cite{Cirelli:2012ut} (Beyesian
versus frequentist), our results are in good agreement with theirs,
proving the consistency of both the approaches.\footnote{The result
  for $\tau^+\tau^-$ in high mass region is different from
  Ref.\,\cite{Cirelli:2012ut}. This is because they used both
  published and preliminary data for $E_\gamma>100~{\rm GeV}$ (at that
  time) while we use the published 10-month data only. In
  $\tau^+\tau^-$ case, gamma-ray spectrum from cascade decay is hard
  and the peak of the intensity is out of data region when $m_{\rm
    dm}\gtrsim ~{\rm TeV}$. Then the constraint becomes weaker.  } In
Fig.~\ref{fig:limit_tau_PL}, we show results by using the
phenomenological cutoff power-law model as the astrophysical
component.  Again, this is just for reference purpose, since such a
single-component astrophysical modeling is no longer valid.

Finally in Fig.~\ref{fig:limit_DMonly} we repeated our analysis by
switching off all astrophysical components, i.e.  taking into account
only dark matter component, for reference. This result corresponds to
the most conservative limit on the lifetime, which may be helpful for
some readers.

\subsection{Implications for particle physics models}

As discussed, dark matter models in which large amount of high-energy
electrons and positrons are produced are phenomenologically motivated
in order to explain the observed cosmic-ray positron excess. Typically
a lifetime of $10^{26}~{\rm s}$ with a mass of $\sim$1~TeV is required
to account for the excess. Our result indicates that such dark matter
models are almost excluded already.

The parameter region of decaying gravitino via $LH_u$ RPV,
$m_{3/2}=1$--2~TeV and $\tau_{3/2}\simeq 10^{26}~{\rm
  s}$~\cite{Carquin:2015uma} to explain the latest AMS-02 positron
data, is excluded in both Normal and Flat priors
(Figs.~\ref{fig:limit_tau_Normal} and \ref{fig:limit_tau_Flat}). In
Ref.\,\cite{Carquin:2015uma} constraint from the EGRB is also
discussed. Taking into account the astrophysical sources and 50-month
Fermi-LAT data, they concluded that the decaying gravitino with the
positron excess-motivated region is not excluded, which is
inconsistent with our result.\footnote{This statement was based on the
  first version of Ref.\,\cite{Carquin:2015uma}. In the second version
  they fixed a bug in the computation of gamma-ray flux and gave a
  consistent result with ours.  We thank German Gomez-Vargas for
  pointing this out.}  The decay channel $\mu^+\mu^-$ to explain the
positron excess is excluded as well.  The same conclusion is already
stated by Ref.\,\cite{Cirelli:2012ut}. We have confirmed their
conclusion in more sophisticated treatment for astrophysical sources.
Lastly decaying wino dark matter via $LLE^c$ interaction to give
sufficient positron flux to explain AMS-02 data is also excluded for
thermal wino with a mass of 3 TeV.

The lifetime of dark matter which decays mainly through $U^cD^cD^c$
RPV is severely constrained too.  In such case, though decaying dark
matter no longer explains the positron excess, decaying axino, for
instance, with a mass of ${\cal O}(10~{\rm GeV})$ is motivated in the
context of baryogenesis, which has been mentioned in the previous
sections.  The results obtained in this study shows that
$\tau_{\tilde{a}}=10^{26}$--$10^{27}~{\rm s}$ for ${\cal O}(10~{\rm
  GeV})$ axino mass are partly excluded.  This means that the
parameter space of axino dark matter in moduli-induced baryogenesis
may be probed in the future observation.

Here we have discussed the implication of our numerical results for
the dark matter models described in Sec.~\ref{sec:DMmodel}. However,
it is obvious that the results given in
Sec.~\ref{sec:ConstraintsOnTauDM} does not depend on what dark matter
is. Only the decay products of dark matter, i.e. the final state,
change the constraints on the lifetime of dark matter (as function of
its mass). Thus, if gravitino decays via $LLE^c$ RPV, for example,
then the result of (a) in Figs.~\ref{fig:limit_tau_Normal} and
\ref{fig:limit_tau_Flat} can be roughly applied.\footnote{Strictly
  speaking, the energy distribution of charged leptons from gravitino
  decaying in three body are different. However, the difference will
  not change the constraints on the lifetime significantly.}

\section{Discussion}
\label{sec:discussion}
\setcounter{equation}{0} 

\subsection{Comparison with constraints from galaxy clusters and dwarf
galaxies}

There are other regions of the sky that can be used to constrain dark
matter decay through gamma rays.  The largest virialized dark matter
structures, known as clusters of galaxies, were studied in the
literature for the purpose~\cite{cluster1, cluster2}.  Dwarf
spheroidal galaxies are nearby dark matter substructure in the Galaxy
that are highly dark matter dominated, and hence they are considered
ideal for dark matter searches.  The dwarf galaxies were also studied
for constraining dark matter decay~\cite{cluster1}.  Here, we compare
them with the EGRB that was used in our study.

Gamma-ray intensity $\Phi_\gamma$ is proportional to line-of-sight
integral of dark matter density as was discussed in
Sec.~\ref{sec:formulation}.  In case of the EGRB where the integration
should be taken up to the Hubble horizon, this quantity is
\begin{equation}
 J_{\rm EGRB} = \frac{c \Omega_{\rm dm} \rho_c}{H_0}
  \int dz \frac{1}{\sqrt{\Omega_\Lambda + \Omega_{\rm
  m}(1+z)^3}}\ ,
\end{equation}
and is $3\times 10^{22}$~GeV\,cm$^{-2}$ if we integrate over $0 < z <
5$.  Note that we here ignored gamma-ray absorption as well as the
gamma-ray spectrum per decay $dN_\gamma / dE$ found in
Eq.~(\ref{eq:intensity}), on order to have order-of-magnitude insight
of the relevant term.  This is now compared with the same quantity for
the galaxy clusters and dwarf galaxies, after smeared over the
Fermi-LAT angular resolution.  Reference~\cite{cluster1} computed it
for both the sources, and found that they were typically $2\times
10^{22}$~GeV\,cm$^{-2}$ for the most promising cluster (Fornax) and
more than one order of magnitude smaller for all the known dwarf
galaxies, after averaging over the region with $1^\circ$ radius.  They
are not only smaller than $J_{\rm EGRB}$ but also subject to
uncertainties on dark matter density profiles.  In addition, the EGRB
spectrum is inferred from the all-sky data, where there are many more
photons available for the analysis than small regions around galaxy
clusters or dwarf galaxies.  Therefore, we conclude that the EGRB
provides the best opportunity to constrain dark matter decay with
gamma rays.

\subsection{Comparison with constraints from Galactic anti-protons}

If dark matter decays hadronically, it produces lots of protons and
anti-protons. Since high-energy anti-protons are rare in canonical
Galactic activities, measurements of cosmic-ray anti-proton constrain
hadronic decay of dark matter. However, it is known that there is huge
theoretical uncertainty in the calculation of cosmic-ray anti-proton
in the Galactic region. Cosmic-ray anti-proton flux can be calculated
in a Galactic propagation model, which has several parameters. They
are determined by using the measurements of other cosmic rays, such as
proton flux and boron-to-carbon ratio, etc., which unfortunately
cannot determine all of the parameters. Propagation models called
`MAX', `MED' and `MIN' are the typical
examples~\cite{Delahaye:2007fr}. Although they are consistent with the
measurements of other light elements in cosmic rays, they give
different predictions for anti-proton flux.  In decaying dark matter
model, the anti-proton flux changes up to an order of magnitude
between MAX and MIN models (see Ref.\,\cite{Cirelli:2013hv}, for
example).\footnote{See also
  Refs.\,\cite{Kappl:2014hha,Boudaud:2014qra} for recent
  development. } In addition, the constraint becomes much weaker in
low dark matter mass region, i.e. $m_{\rm dm}\lesssim 100~{\rm
  GeV}$~\cite{Cirelli:2013hv}. On the contrary, the constraints from
the EGRB we have studied in this paper have less uncertainty compared
to those in cosmic ray $\bar{p}$. It is noticed that there also exists
an uncertainty which is involved in Galactic propagation model. The
50-month data from Fermi-LAT~\cite{Ackermann:2014usa} is given by
using GALPROP package~\cite{galprop} in order to simulate diffuse
gamma rays in the Galactic region. They discuss the uncertainty by
computing inner-Galactic gamma rays in different Galactic models,
which corresponds to models A, B, and C described in the previous
sections.  Our results show that the constraints on dark matter model
changes within an order of magnitude in the three models. Thus
observation of the EGRB gives much more robust limits on decaying dark
matter model with less uncertainty compared to the limits from
anti-proton flux. In addition, it should be noted that the 50-month
data gives much more stringent constraint on the lifetime than
10-month data in all types of dark matter models studied here. Thus
more accumulated data with expanded analyses including
anisotropies~\cite{anisotropy,Camera:2012cj} will give probably the
best limits on decaying dark matter models in the future.

\section{Conclusions}
\label{sec:conclusion}
\setcounter{equation}{0}

We have studied the extragalactic gamma-ray background in decaying
dark matter models.  Decaying dark matter produces gamma rays both
primarily from the cascade decay and from electron/positron through
inverse Compton scattering off the microwave-background photons.  If
the final state, along with its mass and lifetime of decaying dark
matter, is specified, the gamma-ray spectrum can be computed with
little theoretical uncertainty. Astrophysical sources of gamma rays in
extragalactic region, on the other hand, have been resolved with
gamma-ray telescopes such as Fermi. Main components, which are
blazars, star-forming galaxies and misaligned active galactic nuclei,
give rise to consistent flux with the measurement of the gamma-ray
background according to Fermi. We have used the updated astrophysical
models for those components and the latest Fermi data after 50-month
sky survey to constrain decaying dark matter models. In order to make
qualitative statement for various class of models, we have considered
supersymmetric dark matter in $R$-parity violation.  Supersymmetric
model accommodates variety of dark matter candidates, such as wino,
gravitino, sneutrino and axino and so on. They decay to standard-model
particles leptonically or hadronically, depending on the relevant
$R$-parity violating interaction. We have calculated gamma-ray flux in
the models where dark matter mainly decays to $l^+_jl^-_k$, $\nu_i
e_j^-e_k^+$ ($\bar{\nu}_i e_j^+e_k^-$), $W^{\pm}l_i^{\mp}$,
$u_id_jd_k$ ($\bar{u}_i\bar{d}_j\bar{d}_k$), and $q_i\bar{q}_i$. The
numerical analysis shows that the constrains on the lifetime of dark
matter becomes more stringent by a factor of a few to an order of
magnitude, compared to the past work.  As the result, we found that
leptonically (and hadronically) decaying dark matter which explains
the cosmic ray positron excess is excluded in most cases. Hadronically
decaying dark matter is also severely constrained. For example, the
current Fermi data are beginning to exclude part of parameter space of
decaying axino dark matter in moduli-induced baryogenesis, which
indicates that future observation of the extragalactic gamma ray may
probe this model.

\section*{Acknowledgments}

We thank Francesca Calore and Irene Tamborra for providing spectral
data of mAGNs and SFGs, respectively.  SA was supported by the
Netherlands Organization for Scientific Research (NWO) through a Vidi
grant.  KI has been supported in part by the German Science
Foundation (DFG) within the Collaborative Research Center 676
“Particles, Strings and the Early Universe”.



\end{document}